\documentclass[a4paper]{article}
\pdfoutput=1
\usepackage{jcappub}
\usepackage{bbold}
\usepackage{mathtools}
\usepackage{ulem}
\usepackage{float}
\usepackage{enumitem}
\usepackage{diagbox}
\def\laq{~\raise 0.4ex\hbox{$<$}\kern -0.8em\lower 0.62ex\hbox{$\sim$}~}
\def\gaq{~\raise 0.4ex\hbox{$>$}\kern -0.7em\lower 0.62ex\hbox{$\sim$}~}

\newcommand{\class}{{\sc class}}

\def \ra {\rightarrow}

\def \De {\Delta}
\def \de {\delta}

\def \ga {\gamma}

\def\laq{~\raise 0.4ex\hbox{$<$}\kern -0.8em\lower 0.62ex\hbox{$\sim$}~}
\def\gaq{~\raise 0.4ex\hbox{$>$}\kern -0.7em\lower 0.62ex\hbox{$\sim$}~}

\def \ga {\gamma}

\def \ra {\rightarrow}

\newcommand{\bn}{{\bf n}}

\newcommand{\OO}{\mathcal O}
\newcommand{\TT}{\mathcal T}

\newcommand{\GG}{\mathcal G}

\newcommand{\fnl}{f_{\rm nl}}

\def\laq{~\raise 0.4ex\hbox{$<$}\kern -0.8em\lower 0.62ex\hbox{$\sim$}~}
\def\gaq{~\raise 0.4ex\hbox{$>$}\kern -0.7em\lower 0.62ex\hbox{$\sim$}~}

\def\be{\begin{equation}}
\def\ee{\end{equation}}
\def\bea{\begin{eqnarray}}
\def\eea{\end{eqnarray}}
\def\bean{\begin{eqnarray*}}
\def\eean{\end{eqnarray*}}

\title{
The CMB bispectrum from bouncing cosmologies
}

\author{Paola C. M. Delgado$^1$,}
\author{Ruth~Durrer$^2$ and}
\author{Nelson Pinto-Neto$^3$}

\affiliation{$^1$Faculty of Physics, Astronomy and Applied Computer Science, Jagiellonian University,
30-348 Krakow, Poland}
\affiliation{
$^2$Universit\'e de Gen\`eve, D\'epartement de Physique Th\'eorique and CAP,
24 quai Ernest-Ansermet, CH-1211 Gen\`eve 4, Switzerland
}
\affiliation{$^3$Department of Cosmology, Astrophysics and Fundamental Interacions-COSMO, Centro Brasileiro de Pesquisas F\'{\i}sicas-CBPF, rua Dr. Xavier Sigaud 150, 22290-180, Rio de Janeiro, Brazil}

\emailAdd{paola.moreira.delgado@doctoral.uj.edu.pl}
\emailAdd{ruth.durrer@unige.ch}
\emailAdd{nelsonpn@cbpf.br}

\abstract{In this paper we compute the CMB bispectrum for bouncing models motivated by Loop Quantum Cosmology. Despite the fact that the primordial bispectrum of these models is decaying exponentially above a large pivot scale, we find that the cumulative signal-to-noise ratio of the bispectrum induced in the CMB from scales $\ell<30$ is larger than 10 in all cases of interest and therefore can, in principle, be detected in the Planck data.
}

\begin{document}

\maketitle

\section{Introduction}\label{s:intro}
The standard cosmological model, solidly grounded in General Relativity theory (GR) and a variety of cosmological observations (the cosmic microwave background radiation and its anisotropies \cite{cmb}, the abundance of light elements \cite{nucleo}, the features of the distribution of large scale structures and cosmological red-shifts \cite{des,sdss}, among others), asserts that the Universe is expanding from a very hot era dominated by radiation, when the geometry of space was highly homogeneous and isotropic, with very small deviations from this special, symmetric state. However, extrapolating the standard cosmological model back to the past using GR, one necessarily encounters a singularity, where physical quantities diverge. Hence, the model is incomplete: GR is pointing us to its own limits, requiring new physics to understand these extreme situation, which is still under debate.

Assuming that the Universe had a beginning immediately followed by a hot expanding phase implies some important new puzzles, related to initial conditions. The size of regions in causal contact in the Universe is given by the Hubble radius, $R_H\equiv |1/H| \equiv |a/\dot{a}|$, where $a$ is the scale factor, and the overdot represents a derivative with respect to cosmic time. The Hubble radius evolves with respect to the scale factor as

\begin{equation}
\label{RH/a}
\frac{\rm{d}\ln(R_H)}{\rm{d} \ln(a)} = 1-\frac{\ddot{a} a}{\dot{a}^2}.
\end{equation}
If the cosmic fluid has non-negative pressure in the hot era, the Friedmann equations imply $\ddot{a}<0$, so that ${\rm{d}\ln(R_H)}/{\rm{d} \ln(a)}>1$. Hence, in the past of an expanding universe 
the size of cosmological scales we are able to see today, which evolve with the scale factor, were much larger than the Hubble radius. This implies that the basic properties of the cosmic microwave background radiation (CMB), its temperature isotropy and its tiny anisotropies, cannot be explained by causal physics, as the scales presenting these observed properties contained hundreds of causally disconnected regions on the last scattering surface. This is called the `horizon problem'. Furthermore, the observed matter and dark energy density of the Universe today, $\rho_0$, is very close to the total energy density of a Universe with flat  spatial sections, $\rho_c$. Using again the Friedmann equations, the ratio $\Omega (t) \equiv \rho(t)/\rho_c(t)$ evolves as
\begin{equation}
\label{omega}
\frac{{\rm{d}}\mid \Omega (t)-1\mid}{{\rm{d}}t}=-2\frac{\ddot{a}}{\dot{a}^3}.
\end{equation}
As $\dot{a}>0$ and $\ddot{a}<0$, if  $\Omega (t)$ is close to unity today, it must have been much closer to unity at earlier times, a spectacular fine tuning of initial conditions. This is the so called `flatness problem'. 

A simple solution to these puzzles is to evoke that at some early stage a new field dominates the Universe evolution, for which $p/\rho < -1/3$ which implies $\ddot{a}>0$, such that ${\rm{d}\ln(R_H)}/{\rm{d} \ln(a)} < 1$ and 
${\rm{d}}\,| \Omega (t)-1|/{\rm{d}}t<0$. This primordial phase is called inflation \cite{inflation1,inflation2,inflation3,inflation4}, usually driven by a simple scalar field, which can be investigated in the framework of standard quantum field theory in curved spacetime. It not only solves the above puzzles, but it also predicted the observed almost scale invariant marginally red-tilted spectrum of primordial cosmological scalar perturbations~\cite{Mukhanov:1982nu}. Inflation became an essential part of any cosmological model in which the Universe has a beginning.

However, there is an alternative simple solution to the above puzzles if one  assumes that the Universe had a very long contracting decelerating phase before the present expanding era. In this case, Eq.~\eqref{RH/a} implies that the Hubble radius was much bigger than any cosmological scale of physical interest in the far past of the contracting phase, because a contracting universe running backwards in time implies larger scale factors. Also, as $\dot{a}<0$, $\ddot{a}<0$, and $\ddot{a}/\dot{a}^3>0$, Eq.~\eqref{omega} implies that flatness becomes an attractor in the contracting phase, rather than a repeller. Hence, the Universe looks spatially flat to us now because it has not expanded long enough in comparison with the very long contracting phase it experienced in the past.

In realistic models with a contracting phase, there must be a bounce connecting it to the present  expanding phase. In GR  contraction generally leads to the time reversal of the Big Bang singularity, the Big Crunch. To avoid this, bouncing models must necessarily involve new physics. In other words, bouncing models must face the cosmological singularity problem, which is not addressed by inflation. A realistic bouncing model then not only solves the above puzzles related to initial conditions, but it is also complete, i.e., free of singularities. 

The new physics required for bouncing models can be classical extensions of GR \cite{cbounce1,cbounce2,cbounce3,cbounce5,cbounce6,cbounce7,cbounce8,cbounce9}, or quantum gravity effects \cite{qbounce0,qbounce1,qbounce2,qbounce3,qbounce4,qbounce5,qbounce6,qbounce7,qbounce8,qbounce9,qbounce10,qbounce11}. Different models have been investigated in the last decades, and it has been shown that many of them \cite{ob0,ob1,ob2,ob3,ob4} satisfy the constraints imposed by cosmological observations on the properties of primordial cosmological perturbations, and other cosmological features, without the need of an inflationary phase. Nevertheless, bouncing models are not incompatible with inflation. Indeed, in some scenarios, bouncing models lead to the appropriate initial conditions for inflation~\cite{bounce-inflation}.

The next  step in this investigation is to find observable `fingerprints', which indicate the existence of a previous contracting phase and a bounce. In Ref.~\cite{Agullo:2020cvg}, in the context of bouncing models with inflation, it was shown that non-Gaussianities originating from the contracting era before inflation \cite{PhysRevD.97.066021} can substantially alleviate the large scale anomalies detected in the CMB~\cite{anomalies}. The model contains a canonical scalar field with potential $V(\varphi)$, and the scale factor around the bounce is generically parametrized as

\begin{equation}
a(t)=a_b (1+b t^2)^n,
\label{abounce}
\end{equation}
where $t$ is cosmic time, the bounce happens at $t=0$, $a_b$ is the scale factor at the bounce, and $b$ is a constant parametrizing the Ricci scalar at the bounce, $R_b=12 n b$. The parameter $n$ controls the way the model enters and leaves the bouncing phase and starts  classical expansion. For $n\approx 1/6$, the scalar field energy density just after the bounce is concentrated in the kinetic term, this is the case of Loop Quantum Cosmology (LQC) models \cite{qbounce10}), and inflation starts later, while for larger values of $n$ the scalar field potential is already relevant at the bounce, and inflation starts earlier. Hence, the features of this class of bouncing models are controlled by the bounce Ricci scalar $R_b$, and by $n$. 

The initial quantum state for the perturbations is chosen to be the adiabatic (Minkowski) vacuum in the far past of the contracting phase. Therefore quantum state of cosmological perturbations at the onset of inflation deviates from the Bunch-Davies vacuum. In terms of the modes one can write as
\begin{equation}
v_k(\eta)=\alpha_k v_k^{\rm BD}(\eta) + \beta_k v_k^{\ast{\rm BD}}(\eta),
\label{advac}
\end{equation}
where $\eta$ is conformal time, implying that the ratio between the primordial dimensionless power spectrum ${\cal P}_{\cal R}(k)$ and the pure Bunch-Davies primordial dimensionless power spectrum ${\cal P}_{\cal R}^{\rm BD}(k)$ reads

\begin{equation}
\frac{{\cal P}_{\cal R}(k)}{{\cal P}_{\cal R}^{\rm BD}(k)}=\mid\alpha_k + \beta_k\mid^2.
\label{power}
\end{equation}

This class of models has two fundamental scales: the bounce scale given by the comoving wave number $k_b \equiv a_b\sqrt{R_b/6}$, and the (comoving) inflation scale 
$k_i \equiv 2\pi a_i\sqrt{R_i/6}$, where $a_i$ and $R_i$ are the scale factor and Ricci scalar at the beginning of inflation, respectively. As the energy scale of the bounce is larger than the energy scale of inflation, $k_i < k_b$. One has three different regimes for the power spectrum: $k>k_b$, $k_i<k<k_b$, and $k<k_i$ (corresponding to length scales smaller than the bounce length scale, bigger than the bounce length scale but smaller than the inflation length scale, and bigger than the inflation length scale, respectively). Scales smaller than the bounce scale do not feel the bounce, hence they will not deviate from inflationary (Bunch-Davies) results. However, the two other scales are affected by the bounce, leading to different physical effects. Indeed, it is shown in Ref.~\cite{Agullo:2020cvg} that non-Gaussianities arise, correlating super-horizon modes with infrared scales, enhancing the probability of the appearance of CMB anomalies at large scales. If the duration of inflation is very long, such effects are suppressed, yielding the constraint $n < 1/4$ for these effects to be significant. The scales which contribute  most to the non-Gaussianity are in the range $k_i<k<k_b$, which is larger for a bounce closer to the Plank scale. 
Hence, models with bounce phases occurring at  length scales larger than 
the Planck length need a larger $f_{\rm nl}$ parameter to yield the desired effects, 
some of them require  $f_{\rm nl}$ of order $10^4$. As the non-Gaussian correlations obtained 
in Ref.~\cite{Agullo:2020cvg} are restricted mostly to super-horizon modes, 
the authors suggest that these large values of 
$f_{\rm nl}$ should not be directly observed. However, the CMB bispectrum of these models is not calculated, and it is not clear under which conditions  the model satisfies the Planck constraints on it~\cite{Planck-nongaussian}.

The aim of this short paper is to fill this gap. We calculate the bispectrum for the two representative models mostly studied in Ref.~\cite{Agullo:2020cvg}: the Loop Quantum Cosmology case $n=1/6$ \cite{qbounce10}, and the $n=0.21$ case, which seems to best mitigate the CMB anomalies according to Ref.~\cite{Agullo:2020cvg}. For these two cases, we consider the minimum and maximum values of $f_{\rm nl}$ allowed ($3326$ and $8518$ for $n=1/6$, and $959$ and $4372$ for $n=0.21$). We test their viability against Planck measurements, and we calculate the signal-to-noise ratio (SNR) of the bispectrum in order to decide whether it can be measured in a CMB experiment which is cosmic variance limited at low multipoles, $\ell<30$, like the Planck experiment~\cite{Planck:2018nkj}. 

The numerical calculations are presented in Section~2. We find that the bispectrum can indeed be quite large at large scales. However, 
for all values $(\ell_1,\ell_2,\ell_3)$ it remains smaller than cosmic variance. Nevertheless, we find that the cumulative SNR for a cosmic variance limited CMB experiment with 70\% sky coverage
becomes larger than 10 for all models considered here.
In Section~3 we end with a discussion and conclusions.

\section{The bispectrum in a bouncing model}

\subsection{The theoretical expressions}
\begin{figure}
\begin{center}\includegraphics[width=7cm]{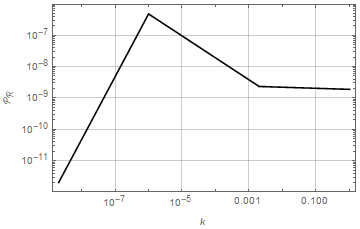}
\end{center}
\caption{\label{f:powerspectrum} 
The power spectrum ${\cal P}_{\cal R}(k)$ versus 
$k$ for $n=1/6$. Note the three different regimes separated by the inflationary and the bounce scales.}
\end{figure}

The bouncing model discussed in~\cite{Agullo:2020cvg} has the following dimensionless power spectrum, ${\cal P}_{\cal R}(k)$, and bispectrum, $B(k_1,k_2,k_3)$, of the curvature fluctuations in Fourier space.

 \bea
{\cal P}_{\cal R}(k) &=& A_s\left\{ \begin{array}{cc} (k/k_i)^2(k_i/k_b)^q & \mbox{if }~k\leq k_i \\
 (k/k_b)^q & \mbox{if }~k_i< k\leq k_b \\
 (k/k_b)^{n_s-1}  &\mbox{if }~k> k_b \,.\end{array}\right. \\
 B(k_1,k_2,k_3) &=& \frac{3}{5}(2\pi^2)^2\fnl\left[\frac{{\cal P}_{\cal R}(k_1)}{k_1^3}\frac{{\cal P}_{\cal R}(k_2)}{k_2^3} +\frac{{\cal P}_{\cal R}(k_1)}{k_1^3}\frac{{\cal P}_{\cal R}(k_3)}{k_3^3}+\frac{{\cal P}_{\cal R}(k_3)}{k_3^3}\frac{{\cal P}_{\cal R}(k_2)}{k_2^3}\right] \times
 \nonumber \\  &&  \qquad \exp\left(-\ga\frac{k_1+k_2+k_3}{k_b}\right)\,. \label{e:Bk}
\eea
Here $n_s=0.9659$ and $A_s=2.3424\times 10^{-9}$, corresponding to the Planck values~\cite{Planck:2018vyg}. The inflation and bounce (pivot) scales are, respectively, $k_i=10^{-6} {\rm Mpc}^{-1}$ and $k_b=0.002 {\rm Mpc}^{-1}$. The parameters $q$, $\fnl$ and $\ga =\sqrt{n \pi/2}\Gamma\left[1-n\right]/\Gamma\left[3/2-n\right]$ depend on the details of the bounce, being related to the parameter $n$ in Eq.~\eqref{abounce}. See \cite{Agullo:2020cvg} for details. The values of $q$ and $\fnl$ used in this work are shown in Table~\ref{table:1}. The power spectrum for $n=1/6$ is shown in Fig.~\ref{f:powerspectrum}.
\begin{table}
\begin{center}
\begin{tabular}{ |c|c|c|c|c| } 
 \hline
 $n$ & $\gamma$ &$q$ & $\fnl$ for $R_{B}=1$ $l_{Pl}^{-2}$ & $\fnl$ for $R_{B}=10^{-3}$ $l_{Pl}^{-2}$ \\ 
 \hline \hline
 $1/6$ &0.6468 & $-0.7$ & $3326$ & $8518$\\
 \hline
 $0.21$ &0.751 &$-1.24$ & $959$ & $4372$\\ 
 \hline
\end{tabular}
\caption{The values of the parameters considered in this work. The $\fnl$ parameter is chosen according to \cite{Agullo:2020cvg} in order to alleviate the power suppression anomaly. The two different values correspond, respectively, to the space-time curvature at the bounce $R_{B}$ equal to $1$ $ l_{Pl}^{-2}$ and to $10^{-3}$ $l_{Pl}^{-2}$, where $l_{Pl}$ is the Planck length.}
\label{table:1}
\end{center}
\end{table}

The bispectrum \eqref{e:Bk} is decaying exponentially for $k>k_b$. Since $k_b$ is close to the horizon scale, the authors of Ref.~\cite{Agullo:2020cvg} argue that the model is not excluded by observations, even for quite large values of $\fnl$. Apart from this exponential decay, which is of course crucial to render a bispectrum with such a large value of $\fnl$ viable, the bispectrum \eqref{e:Bk} is actually just the local bispectrum.
Due to the strong exponential decay, however, its overlap with the local bispectrum is small. In Appendix~\ref{a:shape} we determine numerically the overlap of the LQC bispectrum with the local, equilateral and orthogonal shapes. We find that the overlap with all of them is very small even though the first is somewhat larger than the latter two. The values requested for $\fnl$ given in Table~\ref{table:1} are much larger than the Planck limit for the local shape which is $\fnl\lesssim 10$, see~\cite{Planck:2016aaa}. This is possible since most of the Planck constraint comes from smaller scales, where the LQC bispectrum is exponentially suppressed.

In this work we check quantitatively whether the LQC bispectrum with the parameters given in Table~\ref{table:1} is compatible with observations. Clearly, this non-Gaussianity is  best constrained on very large scales corresponding to $k\lesssim k_b$. This motivates us to  compute the CMB bispectrum induced by it, concentrating on the largest angular scales.
Expanding the CMB temperature fluctuations in spherical harmonics,
\be
\frac{\De T}{T}(\bn) = \sum_{\ell m} a_{\ell m}Y_{\ell m}(\bn) \,,
\ee
the bispectrum is defined by
\be
\langle   a_{\ell_1 m_1} a_{\ell_2 m_2} a_{\ell_2 m_3}\rangle = \GG^{\ell_1\ell_2\ell_3}_{m_1m_2m_3}b_{\ell_1\ell_2\ell_3}  = \left(\begin{array}{ccc}\ell_1&\ell_2&\ell_3\\ m_1&m_2&m_3\end{array}\right)B_{\ell_1\ell_2\ell_3}\,. \label{e:Bisp1}
\ee
Here $\GG^{\ell_1\ell_2\ell_3}_{m_1m_2m_3}$ is the so called `Gaunt factor' which can be expressed in terms of the Wigner $3j$-symbols as
\be
\GG^{\ell_1\ell_2\ell_3}_{m_1m_2m_3} =\sqrt{\frac{\prod_{j=1}^3(2\ell_j+1)}{4\pi}}\left(\begin{array}{ccc}\ell_1&\ell_2&\ell_3\\ 0&0&0\end{array}\right)\left(\begin{array}{ccc}\ell_1&\ell_2&\ell_3\\ m_1&m_2&m_3\end{array}\right)  = g_{\ell_1\ell_2\ell_3}\left(\begin{array}{ccc}\ell_1&\ell_2&\ell_3\\ m_1&m_2&m_3\end{array}\right)\,.
\ee
The $m_i$-dependent prefactor is a consequence of statistical isotropy, see~\cite{Durrer:2020fza} for details. The model dependent quantity $b_{\ell_1\ell_2\ell_3}$ is called the reduced bispectrum. It depends only on the values $\ell_1,\ell_2,\ell_3$ and vanishes if these do not satisfy the triangle inequality, $|\ell_1-\ell_2|\leq \ell_3\leq \ell_1+\ell_2$ or if the sum $\ell_1+\ell_2+\ell_3$ is odd.

Within linear perturbation theory, the  reduced CMB bispectrum is entirely determined by the primordial bispectrum of the curvature fluctuations in Fourier space. More precisely,
\bea
b_{\ell_1\ell_2\ell_3} &=& \left(\frac{2}{\pi}\right)^3\int _0^\infty \!\! \!\! dx\, x^2 \!\int _0^\infty\!\! \!\! dk_1\!\int _0^\infty\!\! \!\! dk_2\!\int _0^\infty\!\! \!\! dk_3 \times \nonumber \\
&& \left[\prod_{j=1}^3\TT(k_j,\ell_j)j_{\ell_j}(k_jx)\right] (k_1k_2k_3)^2B(k_1,k_2,k_3)   \,, \qquad \qquad \label{eNG:Bzeta-blll}
\eea
where $\TT(k,\ell)$ is the CMB transfer function and $j_\ell$ is the spherical Bessel function of index $\ell$, see \cite{Durrer:2020fza} for more details.
On large scales, considering only the  Sachs Wolfe term, we can approximate the transfer function by
\be\label{transfer function}
\TT(k,\ell) \simeq \frac{1}{5} j_\ell(k(t_0-t_{\rm dec})) \,.
\ee
The times $t_0$ and $t_{\rm dec}$ are the (conformal) present time and the decoupling time respectively, given by (we set the speed of light to $c=1$)
\be
t_0\simeq14093.023~{\rm Mpc}\, \qquad t_{\rm dec}\simeq279.529~{\rm Mpc} \,.
\ee
Our bispectrum in $k$-space is separable, i.e., it can be written as a sum of products of functions of $k_i$,
\bea
 (k_1k_2k_3)^2B(k_1,k_2,k_3)  &=& B_0\left[ f(k_1)f(k_2)g(k_3)+ f(k_1)f(k_3)g(k_2)+ f(k_3)f(k_2)g(k_1)\right]
 \qquad  \\  \mbox{ where}  && \nonumber \\
 B_0 &=&  \frac{3}{5}(2\pi^2)^2\fnl  \, \\
 f(k) &=&  \frac{{\cal P}_{\cal R}(k)}{k}\exp(-\ga k/k_b) \qquad \mbox{and} 
 \label{e:fk}\\
 g(k) &=& k^2\exp(-\ga k/k_b)   \label{e:gk}
\eea
Setting 
\bea
X_\ell(x,k) &=& \TT(k,\ell)j_{\ell}(kx)f(k) \qquad \mbox{and} \\
Z_\ell(x,k) &=& \TT(k,\ell)j_{\ell}(kx)g(k)
\eea
with Eq.~\eqref{eNG:Bzeta-blll} we obtain
\bea
b_{\ell_1\ell_2\ell_3} &=& \left(\frac{2}{\pi}\right)^3B_0\int _0^\infty \!\! \!\! dx\, x^2 \!\int _0^\infty\!\! \!\! dk_1\!\int _0^\infty\!\! \!\! dk_2\!\int _0^\infty\!\! \!\! dk_3\Big[ X_{\ell_1}(x,k_1) X_{\ell_2}(x,k_2) Z_{\ell_3}(x,k_3) +  \qquad \qquad \nonumber \\
&&\qquad  X_{\ell_1}(x,k_1) X_{\ell_3}(x,k_3) Z_{\ell_2}(x,k_2) +X_{\ell_3}(x,k_3) X_{\ell_2}(x,k_2) Z_{\ell_1}(x,k_1) \Big]
\eea
This is the sum of three separable $k$-integrals. We introduce 
\bea \label{e:Xell}
X_\ell(x) &=& \int_0^\infty dkX_\ell(x,k)  \qquad \mbox{and}\\
Z_\ell(x) &=& \int_0^\infty dkZ_\ell(x,k)   \label{e:Zell}
\eea
such that the bispectrum becomes the following integral:
\bea\label{bispectrum}
b_{\ell_1\ell_2\ell_3} &=& \left(\frac{2}{\pi}\right)^3B_0\int _0^\infty \!\! \!\! dx\, x^2 \Big[X_{\ell_1}(x)
X_{\ell_2}(x)Z_{\ell_3}(x) +X_{\ell_1}(x)X_{\ell_3}(x)Z_{\ell_2}(x) + \qquad \nonumber \\
&&  \qquad\qquad\qquad \qquad\qquad\qquad  +X_{\ell_3}(x)X_{\ell_2}(x)Z_{\ell_1}(x) \big] \,.
\eea

\subsection{Numerical calculations}

\begin{figure}
\begin{center}\includegraphics[width=6.5cm]{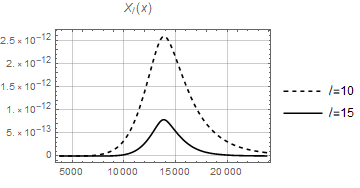}\includegraphics[width=7cm]{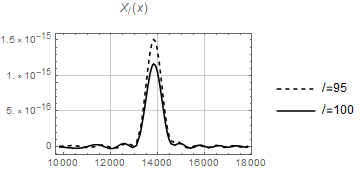}\\
\includegraphics[width=6.5cm]{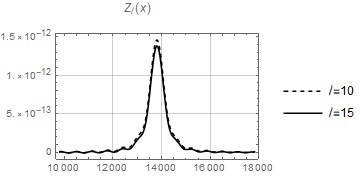}\includegraphics[width=7cm]{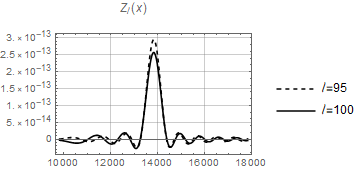}
\end{center}
\caption{\label{f:XlZl} The functions $X_{\ell}(x)$ and $Z_{\ell}(x)$ for $n=1/6$ with the integration over $k$ performed up to $k_{\rm max}=10^{-2}$ ${\rm Mpc}^{-1}$. Note the sharp peaks at $x=t_0-t_{\rm dec}$ and the oscillatory behavior that is especially visible for higher multipoles. Left: $\ell=10$ and $\ell=15$; right: $\ell=95$ and $\ell=100$.}
\end{figure}

\begin{figure}
\begin{center}\includegraphics[width=6.5cm]{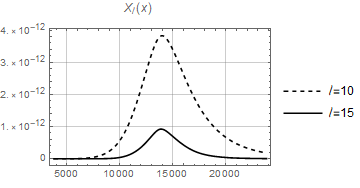}\includegraphics[width=7cm]{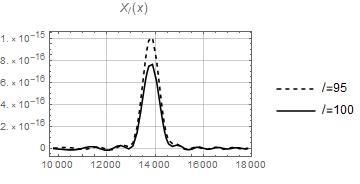}\\
\includegraphics[width=6.5cm]{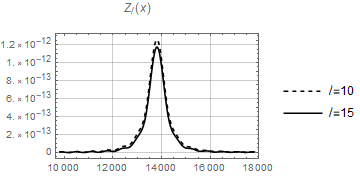}\includegraphics[width=7cm]{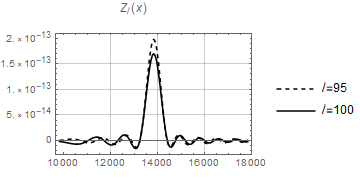}
\end{center}
\caption{\label{f:021XlZl} The functions $X_{\ell}(x)$ and $Z_{\ell}(x)$ for $n=0.21$ with the integration over $k$ performed up to $k_{\rm max}=10^{-2}$ ${\rm Mpc}^{-1}$. Again we see the sharp peaks at $x=t_0-t_{\rm dec}$ and the oscillatory behavior for higher multipoles. Left: $\ell=10$ and $\ell=15$; right: $\ell=95$ and $\ell=100$.}
\end{figure}

The functions $X_\ell(x,k)$ and $Z_\ell(x,k)$ are heavily oscillating as functions of $k$ (containing products of two Bessel functions of different arguments) and difficult to integrate.
However when $x=t_0-t_{\rm dec}$ the product $j_{\ell}(k(t_0-t_{\rm dec}))j_{\ell}(kx)$ becomes a square and both $X_\ell(x,k)$ and $Z_\ell(x,k)$ are positive definite functions of $k$ for this value of $x$. We therefore expect that the integrals over $k$, $X_\ell(x)$ and $Z_\ell(x)$  peak at  $x=t_0-t_{\rm dec}$. As an example, see the functions $X_{\ell}(x)$ and $Z_{\ell}(x)$ in Fig.~\ref{f:XlZl} for $n=1/6$ and in Fig.~\ref{f:021XlZl} for $n=0.21$.
We see that both functions decay rapidly for growing $|x-(t_0-t_{\rm dec})|$ for the two values of $n$ considered.
One might hope, that due to this feature which is the basic idea behind the Limber approximation, the Limber approximation might be relatively good. However, as we show in Appendix~\ref{a:Limber} this is not the case. The Limber approximation actually overestimates the signal by more than one order of magnitude at low $\ell$.

\begin{figure}
\begin{center}\includegraphics[width=7cm]{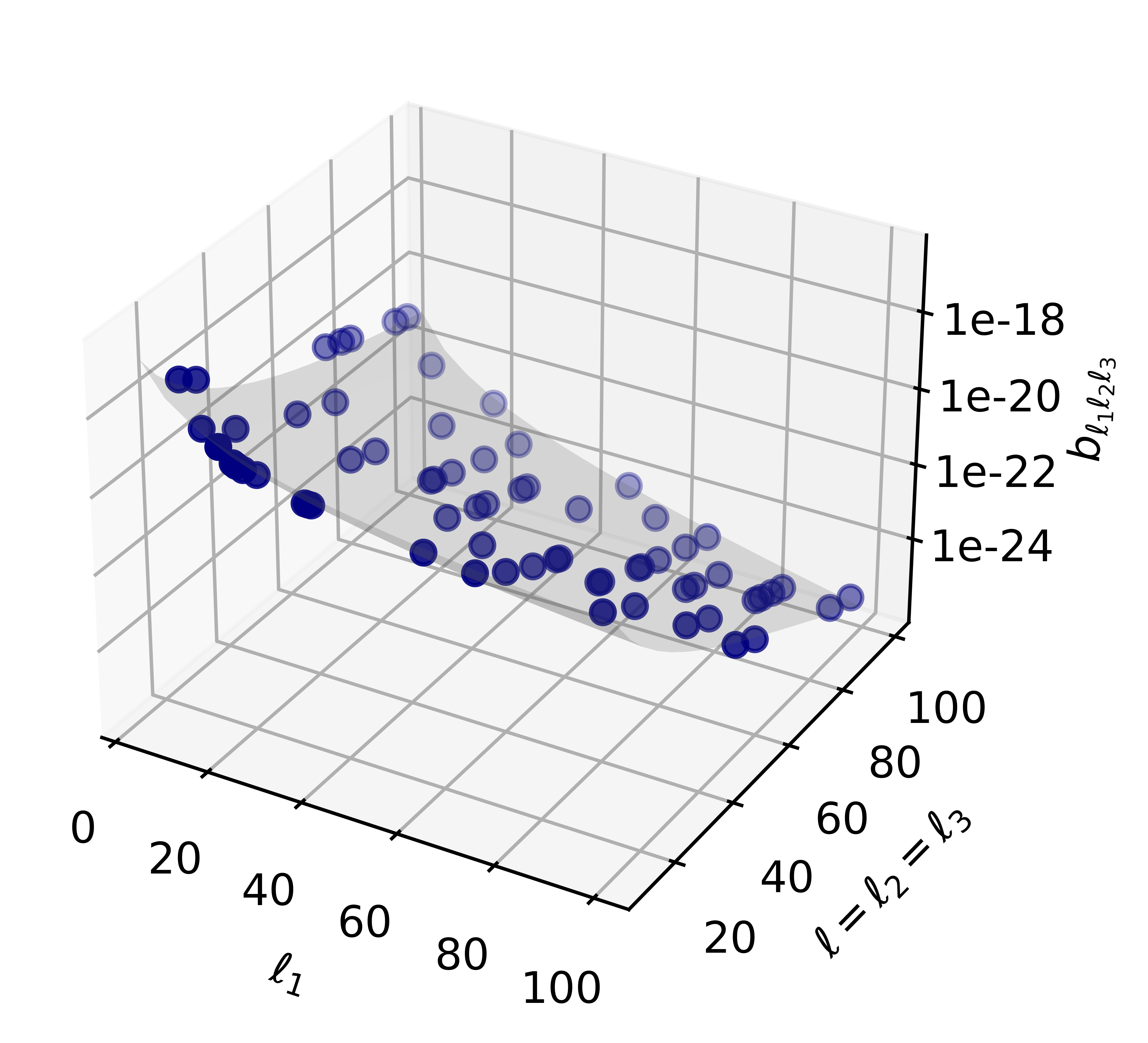}\includegraphics[width=7cm]{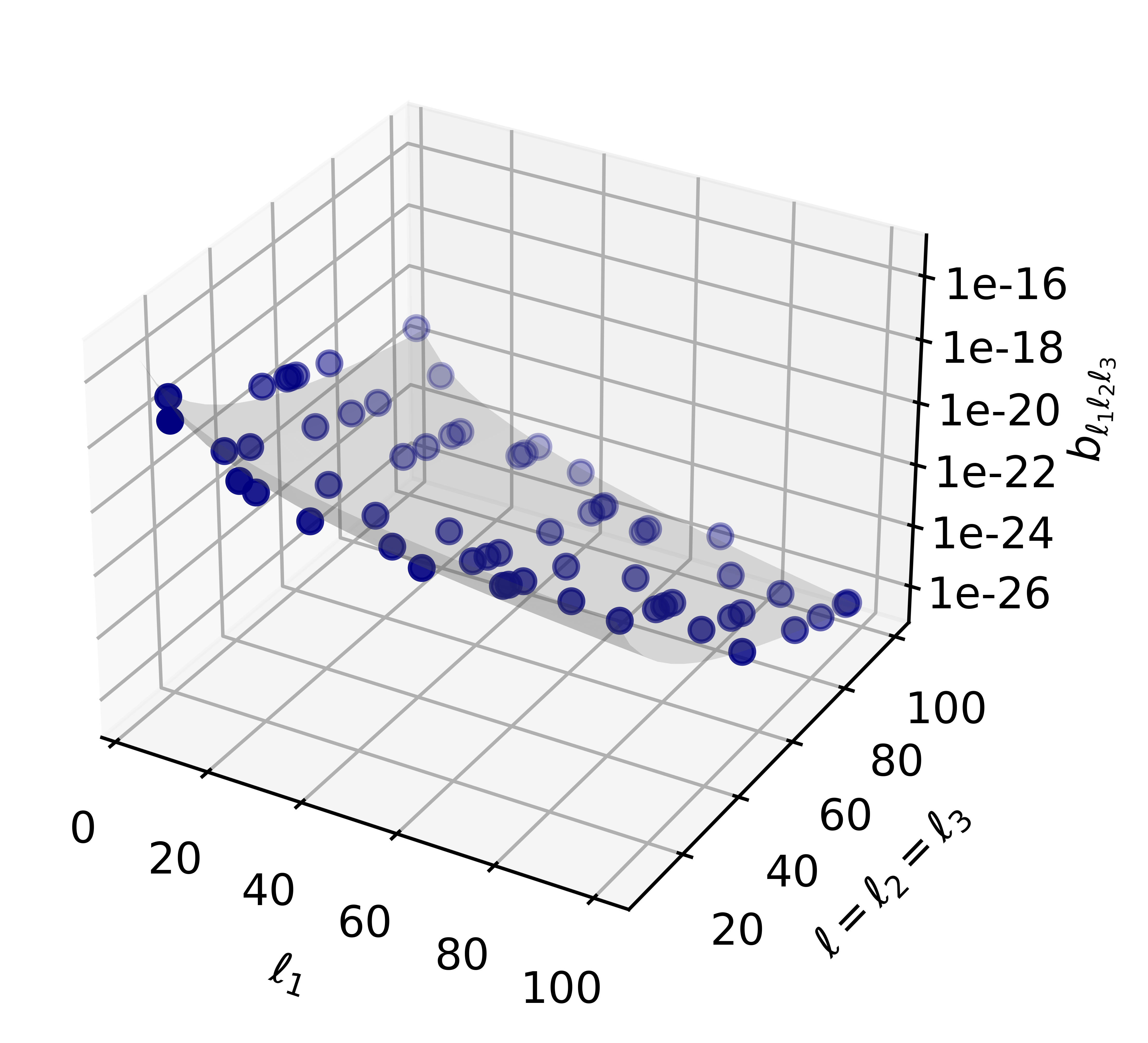}
\end{center}
\caption{\label{f:b3dfnl2} The bispectrum $b_{\ell_1 \ell_2 \ell_3}$ as a function of $\ell_1$ and  $\ell=\ell_2=\ell_3$ for $n=1/6$ (left) and $n=0.21$ (right). The dots correspond to numerical results for $\fnl=8518$ in the case $n=1/6$ and $\fnl=4372$ in the case $n=0.21$. For the other values of $\fnl$ in Table~\ref{table:1}, the plots are re-scaled by the ratios of the $\fnl$'s, the factors of $0.390$ and $0.219$ for $n=1/6$ and $n=0.21$, respectively. The gray planes correspond to the product fits obtained in Section~\ref{signatonoise}, Eqs.~\eqref{fit2} and \eqref{fit4}.}
\end{figure}

The bispectrum \eqref{bispectrum} can be computed numerically if one takes into account the peaked behavior described above, restricting the integration range for the integral over $x$ to an interval around $x=t_0-t_{\rm dec}$. As we can see in Figs.~\ref{f:XlZl} and \ref{f:021XlZl}, the width of the central peak is larger for low values of the multipole $\ell$. For this reason we choose an $x$-range around $t_0-t_{\rm dec}$ of  $10^4$ Mpc for $\ell_1+\ell_2+\ell_3\leq 90$ and $2000$ Mpc for $\ell_1+\ell_2+\ell_3> 90$, encompassing a large  percentage of the total contribution. For the latter cases, the difference with respect to the width $10^4$ is less than $1\%$. Also for the low $\ell$'s the difference between the ranges $10^4$ and $1.5\times 10^4$ is always less than 1\%.  We first perform the computation for a number of allowed sets of multipoles, i.e. with an even sum and satisfying the triangle inequality, starting from the value $\ell_j=4$, $j=1,2,3$, and such that $\ell_2=\ell_3=\ell$. This is a suitable choice in order to depict the bispectrum in a three-dimensional plot. Configurations with $\ell_2\neq \ell_3$ are considered in Section~\ref{signatonoise}. 
The results are plotted in Fig.~\ref{f:b3dfnl2} both for $n=1/6$ (left) and $n=0.21$ (right), where we identify a fast decaying behavior as the multipoles increase. The dots represent the values of the bispectrum obtained via numerical computation for  $\fnl=8518$ in the case $n=1/6$ and $\fnl=4372$ in the case $n=0.21$. The gray planes correspond to the fits obtained in Section~\ref{signatonoise}, namely Eqs.~\eqref{fit2} and \eqref{fit4}.

\begin{figure}
\begin{center}\includegraphics[width=7cm]{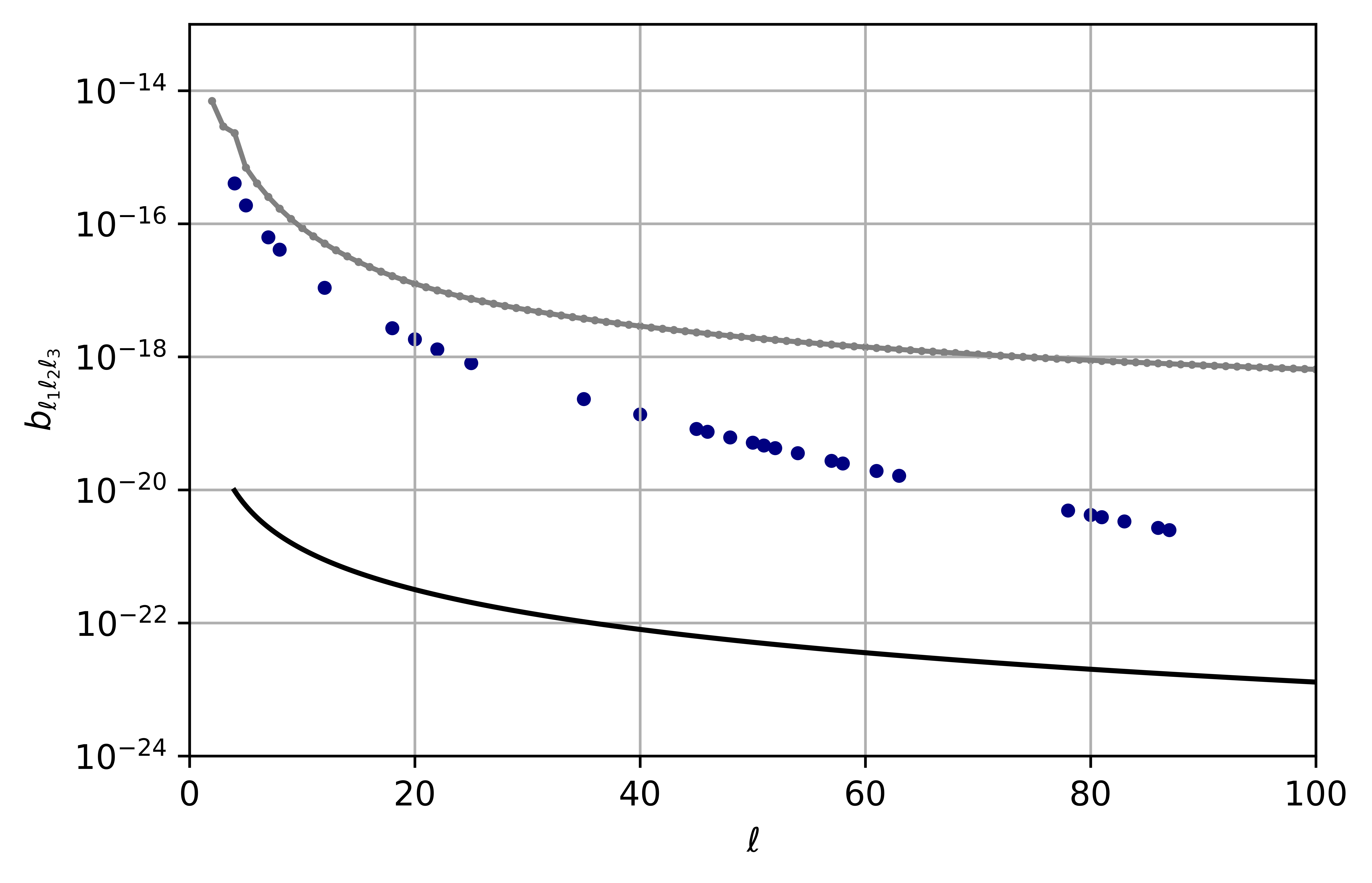}\includegraphics[width=7cm]{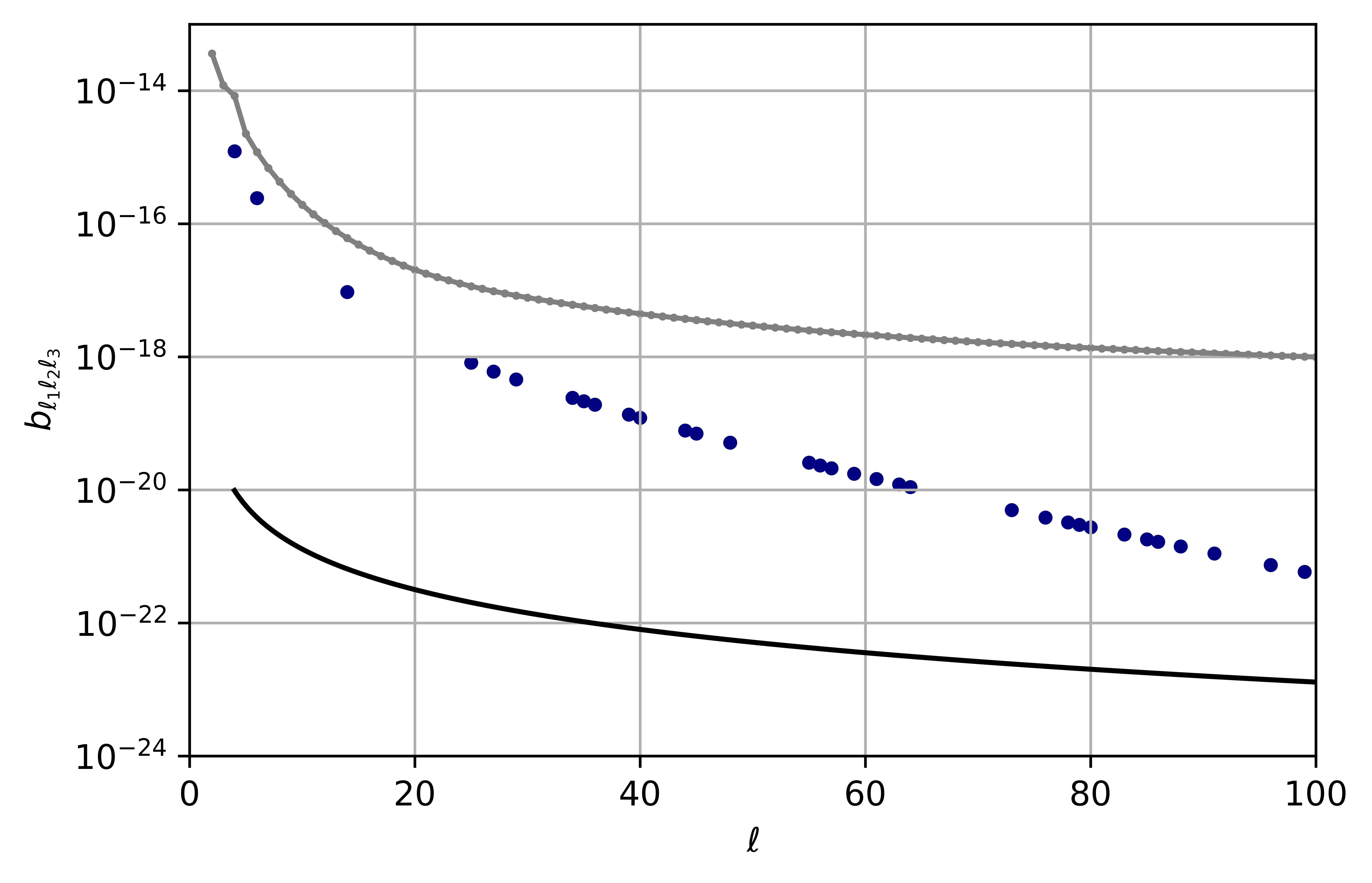}
\end{center}
\caption{\label{f:local} The bispectrum of the current work with $n=1/6$ (left) and $n=0.21$ (right) 
and the local bispectrum with $\fnl=5.0$, considering multipoles such that $\ell_1=4$ and $\ell_2=\ell_3=\ell$. The dots correspond to $\fnl=8518$ for $n=1/6$ and to $\fnl=4372$ for $n=0.21$. For the other values of $\fnl$ in Table~\ref{table:1}, the bispectrum is re-scaled by factors of $0.390$ for $n=1/6$ and $0.219$ for $n=0.21$. The local bispectrum and cosmic variance are depicted as the black and gray lines, respectively.}
\end{figure}

We want to compare the present model and the bispectrum for the local shape. To do so,  we first fix $\ell_1=4$, require $\ell_2=\ell_3=\ell$ and perform the numerical computations for the bispectrum of the current work. 
Then, recalling that the reduced bispectrum of the local shape is given by \cite{Durrer:2020fza}
\begin{equation}
    b_{\ell_1 \ell_2 \ell_3}^{(\operatorname{local})}=\frac{3 \fnl (2\pi^2 A_s)^2}{4\times 5^4}\left( \frac{1}{\ell_1(\ell_1+1)\ell_2(\ell_2+1)}+\frac{1}{\ell_1(\ell_1+1)\ell_3(\ell_3+1)}+\frac{1}{\ell_2(\ell_2+1)\ell_3(\ell_3+1)} \right),
\end{equation}
we substitute $\ell_1=4$, $\ell_2=\ell_3=\ell$ and $\fnl=5.0$, this value of $\fnl$ is chosen based on the Planck constraint on local non-Gaussianity~\cite{Planck-nongaussian}. In Fig.~\ref{f:local}, we plot the power spectra versus $\ell$. 

The bispectrum of the bounce followed by an inflationary phase is larger than the local bispectrum for all the low multipoles considered here. However, this does not mean that it is ruled out by the Planck observations, as most of the observational power from Planck limiting the bispectrum comes from higher values of $\ell$ which are not present in this plot and for which $b^{(\operatorname{local})}$ is much larger than the bispectrum from our bouncing models.

\subsection{Cosmic variance}

In order to decide whether the bispectrum of the bouncing 
models discussed here can be ruled out, we
 consider its signal-to-noise by adding cosmic variance as the dominant noise source on large scales.
This is the minimal error on $b_{\ell_1\ell_2\ell_3}$. Here we follow~\cite{DiDio:2018unb}.
Let us introduce the random variable
\be
\hat B_{\ell_1\ell_2\ell_3} = \sum_{m_1m_2m_3}\left(\begin{array}{ccc}\ell_1&\ell_2&\ell_3\\ m_1&m_2&m_3\end{array}\right) a_{\ell_1 m_1} a_{\ell_2 m_2} a_{\ell_2 m_3}\,.
\ee
This is an estimator of $B_{\ell_1\ell_2\ell_3}$ defined in \eqref{e:Bisp1}, i.e. $\langle \hat B_{\ell_1\ell_2\ell_3}\rangle = B_{\ell_1\ell_2\ell_3}$. In the same way as
\be
\hat C_\ell = (2\ell+1)^{-1}\sum_m|a_{\ell m}|^2
\ee
is an estimator of the power spectrum $C_\ell = \langle |a_{\ell m}|^2\rangle$.
Using identities of the Wigner 3j symbols and neglecting terms involving the bispectrum, which are much smaller than the terms from the power spectrum to the third power, one finds for the variance of our estimator $ \hat B_{\ell_1\ell_2\ell_3}$
\be
{\rm var} (B_{\ell_1\ell_2\ell_3}) = \langle \hat B_{\ell_1\ell_2\ell_3}^2\rangle \simeq C_{\ell_1}C_{\ell_2}C_{\ell_3}\left(1 +\de_{\ell_1\ell_2}  +\de_{\ell_1\ell_3} +\de_{\ell_3\ell_2} +2\de_{\ell_1\ell_2}\de_{\ell_2\ell_3}\right)\,.
\ee
For the reduced bispectrum this yields
\be\label{variance}
{\rm var}\left(b_{\ell_1\ell_2\ell_3}\right)   \simeq g^{-2}_{\ell_1\ell_2\ell_3}C_{\ell_1}C_{\ell_2}C_{\ell_3}\left(1 +\de_{\ell_1\ell_2}  +\de_{\ell_1\ell_3} +\de_{\ell_3\ell_2} +2\de_{\ell_1\ell_2}\de_{\ell_2\ell_3}\right)\,.
\ee
Of course this equality is only valid when $g_{\ell_1\ell_2\ell_3}\neq 0$, i.e., for values of $\ell_1,\ell_2,\ell_3$ which satisfy the triangle inequality and are such that $\ell_1+\ell_2+\ell_3$ is even, since otherwise $b_{\ell_1\ell_2\ell_3} =0$ with vanishing variance.

The minimal error on $b_{\ell_1\ell_2\ell_3}$ an experiment measuring all $ a_{\ell_1 m_1} a_{\ell_2 m_2} a_{\ell_2 m_3}$ with negligible instrumental noise is (the square root of) the cosmic variance. The latter is computed  using the values of $C_\ell$ obtained with the Cosmic Linear Anisotropy Solving System (\class)~\cite{Lesgourgues:2011re,Blas:2011rf} and compared to the amplitude of the bispectrum of the bouncing model in Fig.~\ref{f:local}. Clearly, the amplitude of the square root of the cosmic variance is larger than the bispectrum for all values of $\ell$. This precludes a measurement of the individual $b_{\ell_1\ell_2\ell_2}$'s, but in order to investigate whether  the model can be ruled out due to its non-Gaussianity, we have to go on and compute the total signal-to-noise ratio.

\subsection{Signal-to-noise ratio}\label{signatonoise}
For each individual triple $(\ell_1,\ell_2,\ell_3)$ with $\ell_i\geqslant 4$, $i=1,2,3$, cosmic variance is larger than the value of the bispectrum. However, this does not mean that such a bispectrum is not detectable. To decide on that, we estimate the cumulative signal-to-noise ratio (SNR) of the entire bispectrum for $\ell_i\leq \ell_{\max}$. We choose $\ell_{\max}=80$ to make sure that the Sachs-Wolfe term calculated here really is the dominant contribution. However, as we shall see, the SNR saturates already at $\ell_{\max}\sim 30$.
\be
\left(\frac{S}{N}\right)^2(\ell_{\max}) = \sum_{\ell_1\ell_2\ell_3 =2}^{\ell_{\max}} \frac{b_{\ell_1\ell_2\ell_3}^2}{{\rm var}\left(b_{\ell_1\ell_2\ell_3}\right)} \,.
\ee

\begin{figure}
\begin{center}\includegraphics[width=7cm]{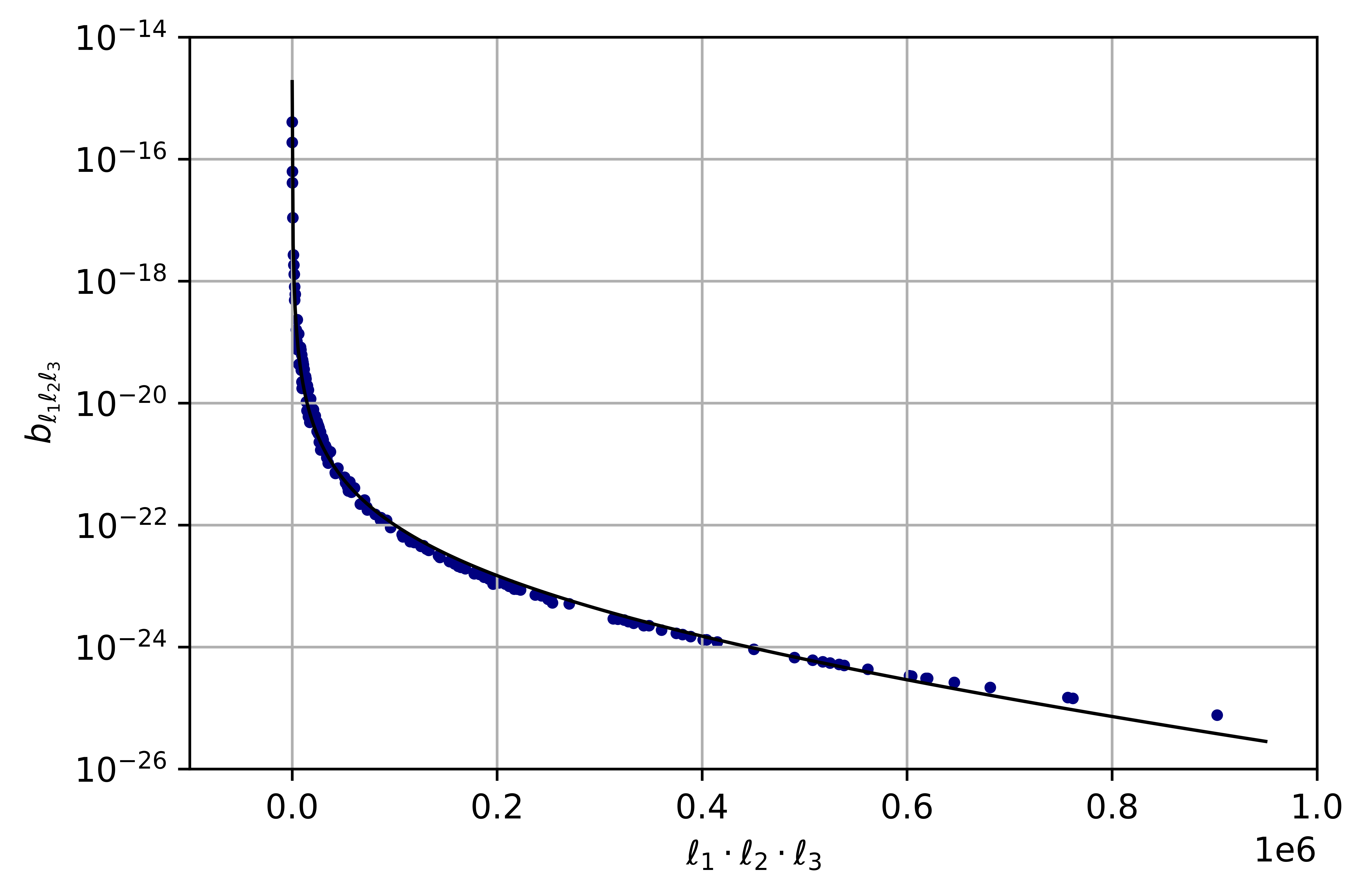}\includegraphics[width=7cm]{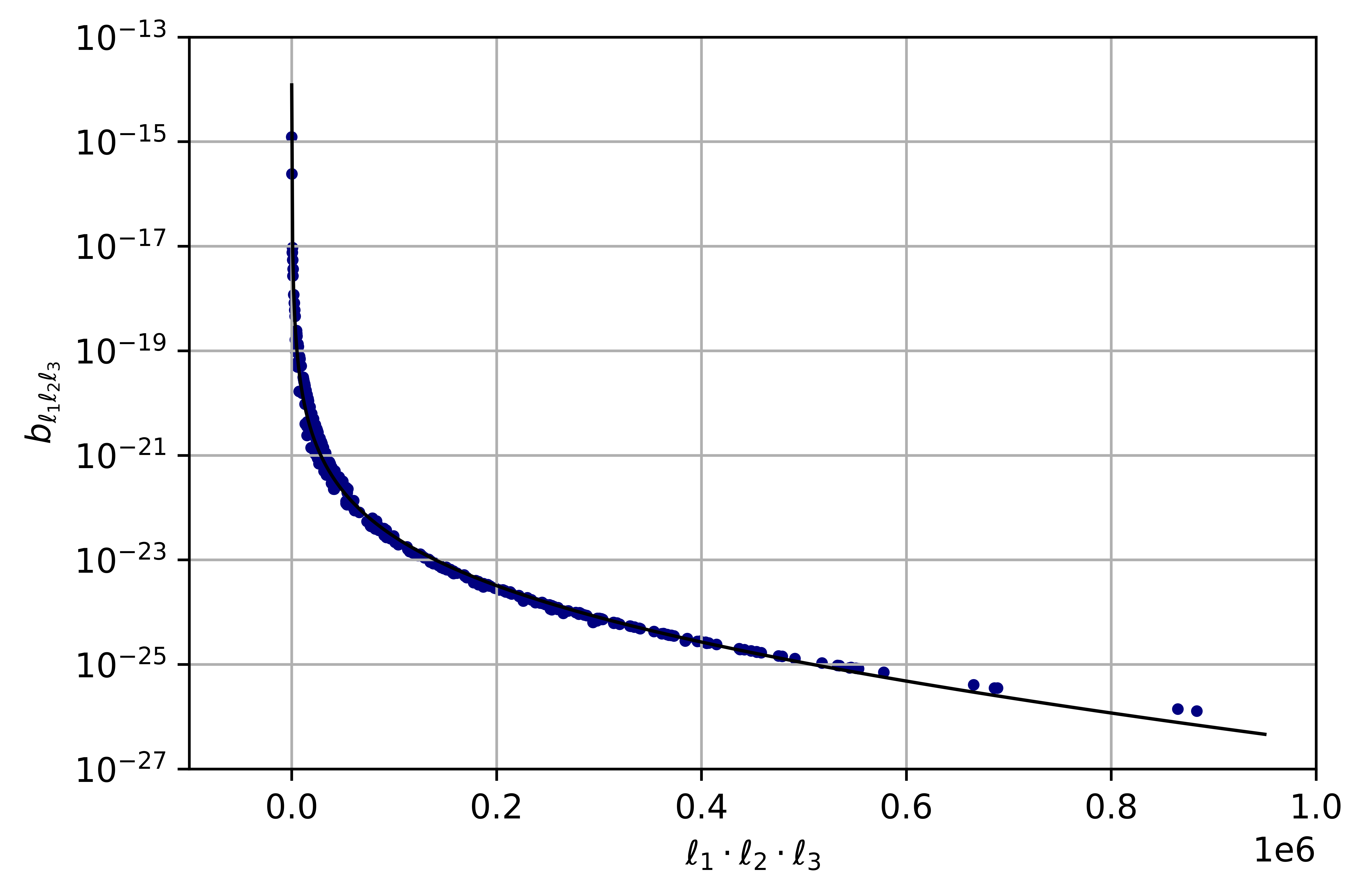}
\end{center}
\caption{\label{f:bvsprod} The bispectrum $b_{\ell_1 \ell_2 \ell_3}$ vs $\ell_1\ell_2\ell_3$ for $n=1/6$ (left) and $n=0.21$ (right). The dots correspond to $\fnl=8518$ for $n=1/6$ and to $\fnl=4372$ for $n=0.21$. The lines represent the fits, given by Eqs.~\eqref{fit2} and \eqref{fit4}. For the other values of $\fnl$ in Table~\ref{table:1}, the approximations are simply re-scaled by the ratios of the $\fnl$ values, namely the factors  $0.390$ for $n=1/6$ and $0.219$ for $n=0.21$.}
\end{figure}

In order to perform this computation, we fit the numerical results of the bispectrum obtained for different sets of multipoles by a product ansatz, including the ones where $\ell_1\neq \ell_2\neq \ell_3$. The fits of our product approximation read
\begin{eqnarray}\label{fit1}
    \operatorname{ln}(b_{\ell_1 \ell_2 \ell_3})&=&-3.727\times10^{-6}(\ell_1 \ell_2 \ell_3)  -2.225 \operatorname{ln}(\ell_1 \ell_2 \ell_3) -25.607
\end{eqnarray}
for $n=1/6$ and $\fnl=3326$,
\begin{eqnarray}\label{fit2}
    \operatorname{ln}(b_{\ell_1 \ell_2 \ell_3})&=&-3.727\times10^{-6}(\ell_1 \ell_2 \ell_3)  -2.225 \operatorname{ln}(\ell_1 \ell_2 \ell_3) -24.667 
\end{eqnarray}
for $n=1/6$ and $\fnl=8518$,
\begin{eqnarray}\label{fit3}
    \operatorname{ln}(b_{\ell_1 \ell_2 \ell_3})=-3.204\times10^{-6}(\ell_1 \ell_2 \ell_3)  -2.661 \operatorname{ln}(\ell_1 \ell_2 \ell_3) -22.491
\end{eqnarray}
for $n=0.21$ and $\fnl=959$ and 
\begin{eqnarray}\label{fit4}
    \operatorname{ln}(b_{\ell_1 \ell_2 \ell_3})=-3.204\times10^{-6}(\ell_1 \ell_2 \ell_3)  -2.661 \operatorname{ln}(\ell_1 \ell_2 \ell_3) -20.974
\end{eqnarray}
for $n=0.21$ and $\fnl=4372$. Two of them are shown in Fig.~\ref{f:bvsprod}, while the other two are simply re-scaled by the ratios of the $\fnl$ values. 

The variance is obtained from Eq.~\eqref{variance}, as before. Summing all the terms corresponding to allowed sets of multipoles, i.e. with $\ell_1+\ell_2+\ell_3$ even and $|\ell_1-\ell_2|\leq \ell_3 \leq \ell_1+\ell_2$, we obtain the cummulative SNR  as a function of the  maximum multipoles $\ell_{\rm max}$.

\begin{figure}
\begin{center}\includegraphics[width=7cm]{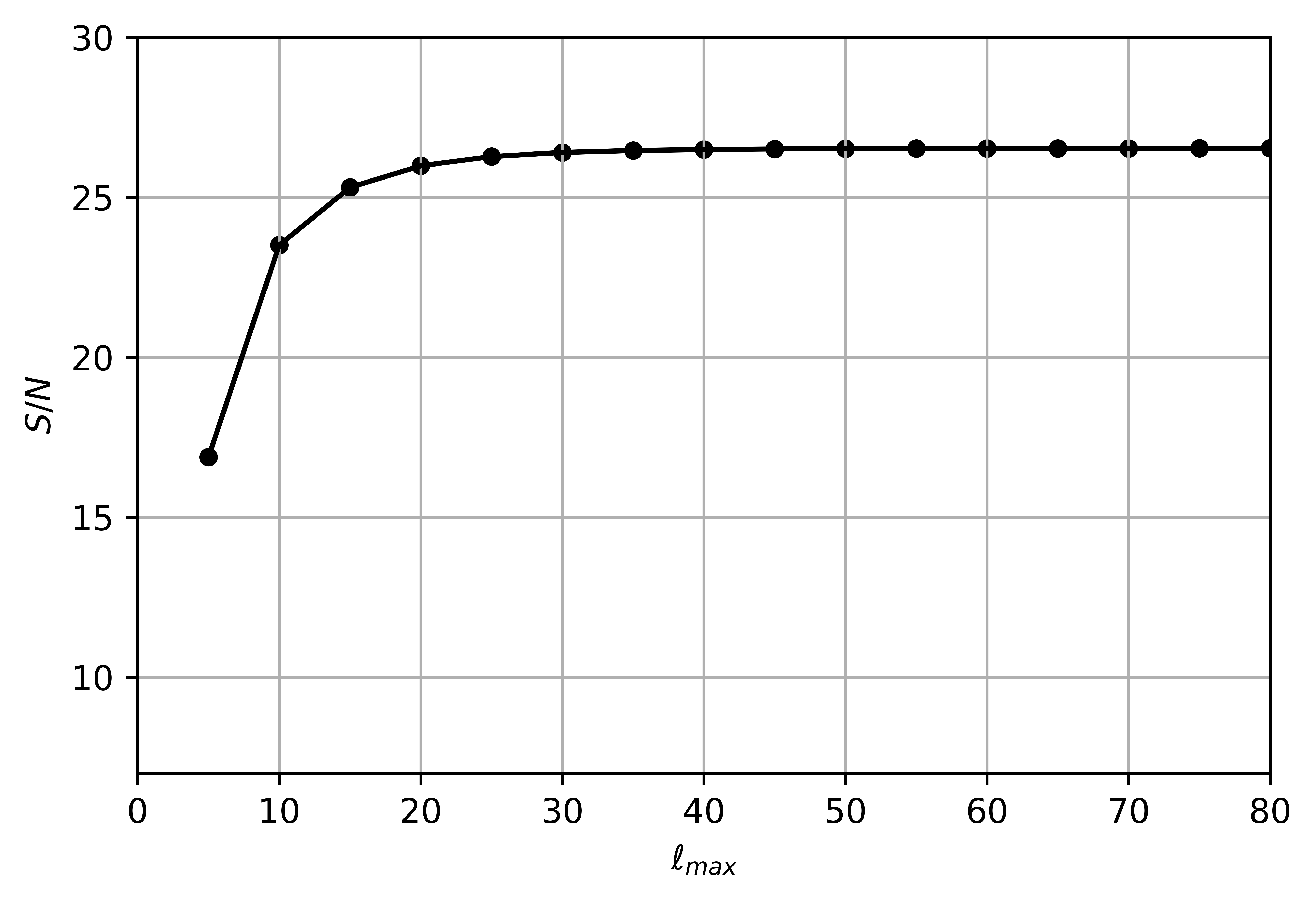}\includegraphics[width=7cm]{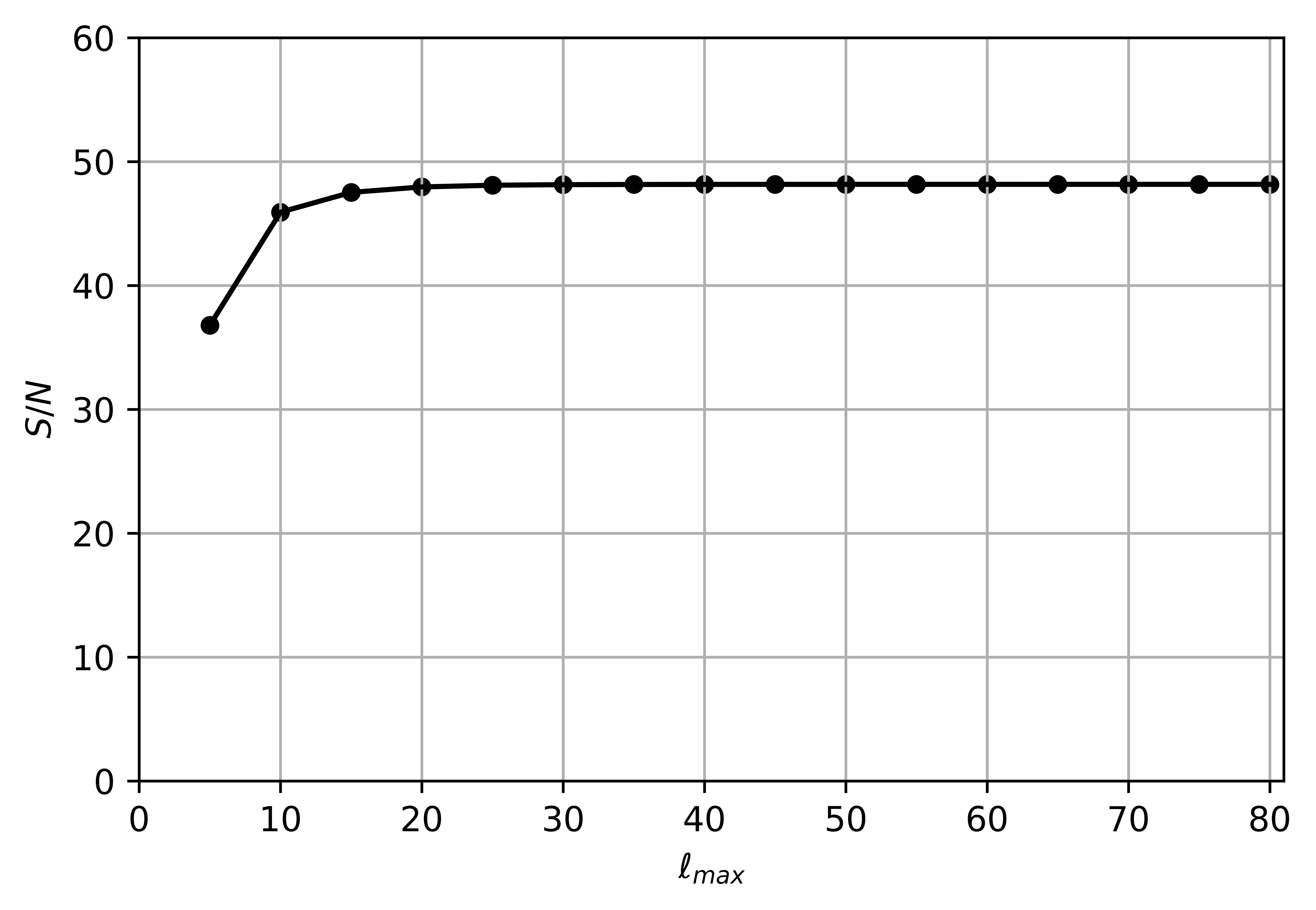}
\end{center}
\caption{\label{f:signaltonoiseproduct} The signal-to-noise ratio, considering $70\%$ of sky coverage, versus maximum values of the multipoles $\ell_{\rm max}$ for $n=1/6$ (left) and $n=0.21$ (right). The dots correspond to $\fnl=8518$ for $n=1/6$ and to $\fnl=4372$ for $n=0.21$. For the other values of $\fnl$ in Table~\ref{table:1}, the bispectrum is re-scaled by factors of $0.390$ for $n=1/6$ and $0.219$ for $n=0.21$.}
\end{figure}

Our purpose it to investigate whether we achieve a value of order $\OO(10)$ within the low $\ell$ regime, i.e. $\ell \ll 200$, which corresponds to the validity of the transfer function given in Eq.~\eqref{transfer function}, and which is also the regime in which the non-Gaussianitiy of these models is larger. The results are shown in Fig.~\ref{f:signaltonoiseproduct}. In order to consider a sky coverage of $70\%$, we multiply the cosmic variance by $1/0.70$. Clearly, the SNR saturates very fast, namely roughly at $\ell_{\max}=30$, but it achieves a value larger than 25, both for $n=1/6$ and $n=0.21$ for the larger value of $\fnl$. For the smaller values of $\fnl$ the cumulative SNR for $n=1/6$ is $10.3$ while for $n=0.21$ it is $10.6$. In all cases these bispectra should be detectable in the Planck data.  

\section{Conclusions}
In this paper we have investigated the CMB bispectrum induced by bouncing models which are motivated by Loop Quantum Cosmology. These models have been proposed in order to address the large scale anomalies of the CMB data. The hope was, that
due to the exponential decay of the bispectrum on subhorizon scales, this non-Gaussianity is not substantial. In this work we show, however, that in all cases with sufficient non-Gaussianity to mitigate the large scale anomalies of CMB data, the bispectrum should be detectable in the Planck data.  Even though the individual $b_{\ell_1\ell_2\ell_3}$ are below cosmic variance if $\ell_i\geqslant 4$, $i=1,2,3$, the cumulative SNR of the bispectrum with $\ell_{\max}=30$ is larger than $10$ for all the models proposed, when assuming a sky coverage of 70\% and considering only temperature data. Note that the largest contributions to the SNR come from triples $(\ell_1,\ell_2,\ell_3)$ where at least one multipole is smaller than $4$, for which the signal is larger than or comparable to the square root of the variance. For the higher values of $\fnl$ the cumulative SNR is about 26.5 ($n=1/6$) and 48.2 ($n=0.21$) respectively. To get an impression of the amplitude of the SNR of these models, one may want to compare it to the one of CMB lensing, which is about 40 in the Planck 2015 data, see~\cite{Planck:2015mym}. Note that already at $\ell_{\max}=5$ the SNR is larger than 15  ($n=1/6$) and 35 ($n=0.21$). But also for the two models with the lower value of $\fnl$, the cumulative SNR is actually just slightly above 10, so that the bispectrum can in principle be detected. In order to obtain an undetectable bispectrum one would have to reduce the $\fnl$ by about a factor of 10, so that the cumulative SNR would become of order unity. However, when reducing $\fnl$ to these values, the CMB large scale anomalies can no longer be resolved efficiently by these models and they lose one of their main attractive features. Furthermore, adding polarisation data may well enhance the SNR by about a factor of two.

These findings motivate us to perform a search for this bispectrum in the actual Planck data.

\section*{Acknowledgements}
We thank Ivan Agullo for discussions. PCMD is supported by the grant No. UMO-2018/30/Q/ST9/00795 from the National
Science Centre, Poland. RD is supported by the Swiss National Science Foundation. NPN acknowledges the support of Conselho Nacional de Desenvolvimento Cient\'{\i}fico e Tecnol\'ogico-CNPq of Brazil under grant PQ-IB
309073/2017-0.

\appendix
\section{Limber approximation}\label{a:Limber}

\begin{figure}[h!t]
\begin{center}\includegraphics[width=14cm]{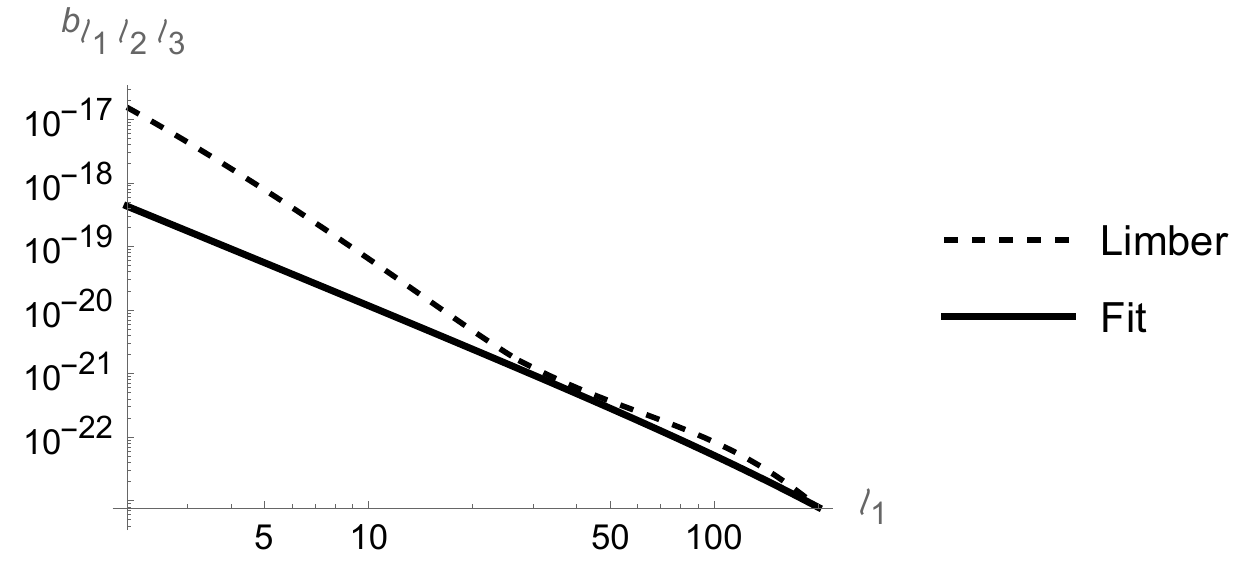}
\end{center}
\caption{\label{f:Limber} The Limber approximation (dashed) is compared with our product approximation (solid) for $\ell_2=\ell_3=30$ fixed, as a function of $\ell_1$. Even though the Limber approximation is probably better for $\ell>50$, it is much worse than our excellent product approximation in the relevant regime, $\ell_i<30$.}
\end{figure}

We also considered to solve the integrals~\eqref{e:Xell} and~\eqref{e:Zell}
with $\TT(k,\ell)=j_\ell(k(t_0-t_{\rm dec}))/5$ using the Limber approximation~\cite{Limber:1953}. This yields
\bea
X_\ell(x) &=& \int dk T(k,\ell)j_\ell(kx)f(k) \simeq \frac{1}{5} \int dk j_\ell(k(t_0-t_{\rm dec}))j_\ell(kx)f(k)\nonumber \\
 &\simeq& \frac{\pi}{10}\frac{\de(t_0-t_{\rm dec}-x)}{(\ell+1/2)^2} f\left(\frac{\ell+1/2}{t_0-t_{\rm dec}}\right)  \nonumber  \\
 &=& \frac{\pi}{10}\de(t_0-t_{\rm dec}-x)\frac{t_0-t_{\rm dec}}{(\ell+1/2)^3}{\cal P}_{\cal R}\left(\frac{\ell+1/2}{t_0-t_{\rm dec}}\right)e^{-\ga\frac{\ell+1/2}{k_b(t_0-t_{\rm dec})}} \label{e:X}\\
Z_\ell(x) &=& \int dk T(k,\ell)j_\ell(k)g(k) \simeq \frac{1}{5} \int dk j_\ell(k(t_0-t_{\rm dec}))j_\ell(kx)g(k) \nonumber \\
 &\simeq& \frac{\pi}{10}\frac{\de(t_0-t_{\rm dec}-x)}{(\ell+1/2)^2} g\left(\frac{\ell+1/2}{t_0-t_{\rm dec}}\right) \nonumber \\ 
 &=& \frac{\pi}{10}\de(t_0-t_{\rm dec}-x)\frac{1}{(t_0-t_{\rm dec})^2}e^{-\ga\frac{\ell+1/2}{k_b(t_0-t_{\rm dec})}} \,. \label{e:Z}
 \eea
 Here we used the functions $f$ and $g$ defined in Eqs.~\eqref{e:fk} and~\eqref{e:gk}.
 Of course, the product of three delta functions cannot be integrated, but replacing them by narrow Gaussians we obtain up to an unknown constant $A$, related to the width of the Gaussian,
 \bea
 b_{\ell_1\ell_2\ell_3)} &=& \left(\frac{2}{\pi}\right)^3B_0\int dx x^2X_{\ell_1}(x)X_{\ell_2}(x)Z_{\ell_3}(x) \nonumber \\
&\sim&  \frac{AB_0}{125}\exp\left(-\ga\frac{\ell_1+\ell_2+\ell_3}{k_b(t_0-t_{\rm dec})}\right)
 \left[\frac{{\cal P}_{\cal R}\left(\frac{\ell_1+1/2}{t_0-t_{\rm dec}}\right)}{(\ell_1+1/2)^3}\frac{{\cal P}_{\cal R}\left(\frac{\ell_2+1/2}{t_0-t_{\rm dec}}\right)}{(\ell_2+1/2)^3} +
 \circlearrowleft \right] \,.
\eea
Here $\circlearrowleft$ indicates that the two permutations $(\ell_1,\ell_2) \ra (\ell_1,\ell_3)$ and $(\ell_1,\ell_2) \ra (\ell_3,\ell_2)$ have to be added.
We have tested this approximation with our numerical computations and found that for the relevant values of $\ell$, $\ell<30$, which contribute most to the SNR, it is much less accurate than the product approximation which we use in the main text, see Fig.~\ref{f:Limber}.

\section{Overlap with standard bispectrum shapes}\label{a:shape} 
In this appendix we determine the overlap of the LQC bispectrum with other standard shapes as they are studied, e.g, in the Planck data~\cite{Planck:2016aaa}. For this purpose we define the dimensionless shape function 
\begin{equation}
    S(k_1,k_2,k_3)=\alpha (k_1,k_2,k_3)^2 B(k_1,k_2,k_3),
\end{equation}
where $\alpha$ is a normalization factor such that $S(k_c,k_c,k_c)=1$ for a characteristic scale $k_c$. For the model analyzed in this work, we choose  $k_c=k_b$ so that 
\begin{equation}
    \alpha=\frac{5}{9}\frac{\operatorname{e}^{3\gamma}}{f_{nl}(2 \pi^2)^2 P_{\cal R}^{2}(k_b)}.
\end{equation}
For the standard cosmological scenario, one usually defines the local, equilateral and orthogonal shapes, which are given respectively by
\begin{eqnarray}
S^{\rm (local)}(k_1,k_2,k_3)&=&\frac{1}{3}\left(\frac{k_1^2}{k_2 k_3}+\frac{k_2^2}{k_1 k_3}+\frac{k_3^2}{k_2 k_1}\right), \\
S^{\rm (equi)}(k_1,k_2,k_3)&=&\frac{(k_1+k_2-k_3)(k_2+k_3-k_1)(k_3+k_1-k_2)}{k_1 k_2 k_3},\\
S^{\rm (ortho)}(k_1,k_2,k_3)&=&\frac{-3k_1^3+3[k_1^2-(k_2-k_3)^2](k_3+k_2)}{k_1 k_2 k_3}+\frac{3k_3^2+k_2(3k_2-8k_3)}{k_2 k_3}.
\end{eqnarray}
Since these shapes are not orthogonal to each other in general\footnote{The orthogonal shape is orthogonal to the equilateral one for a weight function $(k_1 k_2 k_3)^{-1}$. }, one may construct a scalar product 
\begin{equation}
    \left \langle S_1,S_2 \right \rangle = \int_V S_1(k_1,k_2,k_3)S_2(k_1,k_2,k_3)w(k_1,k_2,k_3)dk_1 dk_2 dk_3
\end{equation}
within a region $V$ where $|k_1-k_2|\leq k_3\leq k_1+k_2$, the shape functions $S_1$ and $S_2$ do not diverge and the weight function $w(k_1,k_2,k_3)$ is an arbitrary non-negative function. One can then define an ``angle" $\theta$ between the shapes $S_1$ and $S_2$ by
\be
\cos\theta_{12} =\frac{\langle S_1,S_2 \rangle }{\sqrt{\langle S_1,S_1 \rangle \langle S_2,S_2 \rangle} } \,.
\ee
If this angle is large, $\cos\theta_{12}\ll 1$, there is a small overlap.
 We choose a weight function given by
\bea
w(k_1,k_2,k_3) &=& \left\{ \begin{array}{cc} (k_1 k_2 k_3)^4 & \mbox{if }~10^{-4}\leq k_i\leq 0.04 \\
 0 & \mbox{else }
 \end{array}\right. ,
\eea
where $i=1,2,3$. We first perform the integration over $k_3$ with lower and upper limits given respectively by $|k_1-k_2|$ and $k_1+k_2$. The step function is chosen in order to avoid $k_i=0$, since the shape functions diverge for $k_i\ra 0$, and to consider the interesting range for the bispectrum under investigation, i.e. low values of $k$. The power law was chosen so that the expression to be integrated is regular for all values of $k_i$ considered. Performing the integrations\footnote{The integration limits for $dk_1$ and $dk_2$ are chosen to encompass the non-negligible region after integration over $k_3$.} 
\begin{equation}
    \left \langle S^{\rm (bounce)},S^{(j)} \right \rangle=\int_{10^{-4}}^{4\times 10^{-2}}dk_1\int_{10^{-4}}^{4\times 10^{-2}}dk_2\int_{|k_1-k_2|}^{k_1+k_2}dk_3 S^{\rm (bounce)}S^{(j)} w,
\end{equation}
where $(j)={\rm (local), (equi), (ortho)}$, and normalizing the results, we obtain for $n=1/6$ 
\begin{eqnarray}
\nonumber \cos{\theta}^{\rm (bounce, local)} &=& 2.369\times 10^{-4},\\
\nonumber \cos{\theta}^{\rm (bounce, equi)} &=& 2.364 \times 10^{-4},\\
\cos{\theta}^{\rm (bounce, ortho)} &=& -3.985\times 10^{-5}
\end{eqnarray}
and for $n=0.21$
\begin{eqnarray}
\nonumber \cos{\theta}^{\rm (bounce, local)} &=& 7.117\times 10^{-5},\\
\nonumber \cos{\theta}^{\rm (bounce, equi)} &=& 7.071\times 10^{-5},\\
\cos{\theta}^{\rm (bounce, ortho)} &=& -1.206\times 10^{-5}\,.
\end{eqnarray}

The results do not depend on $\fnl$, as it cancels in the expression of the shape function.
The projection is only slightly larger for the local than for the equilateral shape, but it is very small for all three standard shapes. This is a consequence of the rapid exponential decay of the bouncing bispectrum.

\bibliography{bispec}

\providecommand{\href}[2]{#2}\begingroup\raggedright\begin{thebibliography}{10}

\bibitem{cmb}
N.~Aghanim and et~al., {\it {Planck 2018 results. I. Overview and the
  cosmological legacy of Planck}},  {\em Astron. Astrophys.} {\bf 642} (2020)
  A1, [\href{http://arxiv.org/abs/1807.06205}{{\tt arXiv:1807.06205}}].

\bibitem{nucleo}
R.~Cyburt, B.~Fields, K.~Olive, and T.~Yeh, {\it {Big bang nucleosynthesis:
  Present status}},  {\em Rev. Mod. Phys.} {\bf 88} (2016) 015004,
  [\href{http://arxiv.org/abs/1505.01076}{{\tt arXiv:1505.01076}}].

\bibitem{des}
T.~Abbott and et~al., {\it {Dark Energy Survey Year 1 Results: Cosmological
  constraints from cluster abundances and weak lensing}},  {\em Phys. Rev. D}
  {\bf 102} (2020) 023509, [\href{http://arxiv.org/abs/2002.11124}{{\tt
  arXiv:2002.11124}}].

\bibitem{sdss}
B.~Abolfathi and et~al., {\it {The Fourteenth Data Release of the Sloan Digital
  Sky Survey: First Spectroscopic Data from the Extended Baryon Oscillation
  Spectroscopic Survey and from the Second Phase of the Apache Point
  Observatory Galactic Evolution Experiment}},  {\em Astrophys. J. Suppl. Ser.}
  {\bf 235} (2018) 19, [\href{http://arxiv.org/abs/1707.09322}{{\tt
  arXiv:1707.09322}}].

\bibitem{inflation1}
A.~Starobinsky, {\it {Relict gravitation radiation spectrum and initial state
  of the universe}},  {\em JETP Lett.} {\bf 30} (1979) 682.

\bibitem{inflation2}
A.~Guth, {\it {Inflationary universe: A possible solution to the horizon and
  flatness problems}},  {\em Phys. Rev. D} {\bf 23} (1981) 347.

\bibitem{inflation3}
A.~Linde, {\it {A new inflationary universe scenario: A possible solution of
  the horizon, flatness, homogeneity, isotropy and primordial monopole
  problems}},  {\em Phys. Lett. B} {\bf 108} (1982) 389.

\bibitem{inflation4}
V.~Mukhanov and G.~Chibisov, {\it {Quantum fluctuations and a nonsingular
  universe}},  {\em JETP Lett.} {\bf 33} (1981) 532.

\bibitem{Mukhanov:1982nu}
V.~F. Mukhanov and G.~V. Chibisov, {\it {The Vacuum energy and large scale
  structure of the universe}},  {\em Sov. Phys. JETP} {\bf 56} (1982) 258--265.

\bibitem{cbounce1}
R.~Tolman, {\it {On the theoretical requirements for a periodic behaviour of
  the universe}},  {\em Phys. Rev.} {\bf 38} (1931) 1758.

\bibitem{cbounce2}
G.~Murphy, {\it {Big-bang model without singularities}},  {\em Phys. Rev.} {\bf
  8} (1973) 4231.

\bibitem{cbounce3}
M.~Novello and J.~Salim, {\it {Nonlinear photons in the universe}},  {\em Phys.
  Rev. D} {\bf 20} (1979) 377.

\bibitem{cbounce5}
L.~Allen and D.~Wands, {\it {Cosmological perturbations through a simple
  bounce}},  {\em Phys. Rev. D} {\bf 70} (2004) 063515,
  [\href{http://arxiv.org/abs/astro-ph/0404441}{{\tt astro-ph/0404441}}].

\bibitem{cbounce6}
J.~Fabris, R.~Perez, S.~Bergliaffa, and N.~Pinto-Neto, {\it {Born-Infeld-like
  f(R) gravity}},  {\em Phys. Rev. D} {\bf 86} (2012) 103525,
  [\href{http://arxiv.org/abs/1205.3458}{{\tt arXiv:1205.3458}}].

\bibitem{cbounce7}
Y.~Cai, D.~Easson, and R.~Brandenberger, {\it {Towards a nonsingular bouncing
  cosmology}},  {\em J. Cosmol. Astropart. Phys.} {\bf 8} (2012) 020,
  [\href{http://arxiv.org/abs/1206.2382}{{\tt arXiv:1206.2382}}].

\bibitem{cbounce8}
A.~Ijjas and P.~Steinhardt, {\it {Classically stable nonsingular cosmological
  bounces}},  {\em Phys. Rev. Lett.} {\bf 117} (2016) 121304,
  [\href{http://arxiv.org/abs/1606.08880}{{\tt arXiv:1606.08880}}].

\bibitem{cbounce9}
A.~Ilyas, M.~Zhu, Y.~Zheng, Y.~Cai, and E.~Saridakis, {\it {DHOST bounce}},
  {\em J. Cosmol. Astropart. Phys.} {\bf 9} (2020) 002,
  [\href{http://arxiv.org/abs/2002.08269}{{\tt arXiv:2002.08269}}].

\bibitem{qbounce0}
V.~Melnikov and S.~Orlov, {\it {Nonsingular cosmology as a quantum vacuum
  effect}},  {\em Phys. Lett. A} {\bf 70} (1979) 263.

\bibitem{qbounce1}
M.~Gotay and J.~Demaret, {\it {Quantum cosmological singularities}},  {\em
  Phys. Rev. D} {\bf 28} (1983) 2402.

\bibitem{qbounce2}
F.~Tipler, {\it {Interpreting the wave function of the universe}},  {\em Phys.
  Rep.} {\bf 137} (1986) 231.

\bibitem{qbounce3}
J.~de~Barros, N.~Pinto-Neto, and M.~Sagioro-Leal, {\it {The causal
  interpretation of dust and radiation fluid non-singular quantum
  cosmologies}},  {\em Phys. Lett. A} {\bf 241} (1998) 229,
  [\href{http://arxiv.org/abs/gr-qc/9710084}{{\tt gr-qc/9710084}}].

\bibitem{qbounce4}
J.~Colistete, R., J.~Fabris, and N.~Pinto-Neto, {\it {Gaussian superpositions
  in scalar-tensor quantum cosmological models}},  {\em Phys. Rev. D} {\bf 62}
  (2000) 083507, [\href{http://arxiv.org/abs/gr-qc/0005013}{{\tt
  gr-qc/0005013}}].

\bibitem{qbounce5}
M.~Bojowald, {\it {Absence of a singularity in loop quantum cosmology}},  {\em
  Phys. Rev. Lett.} {\bf 86} (2001) 5227,
  [\href{http://arxiv.org/abs/:gr-qc/0102069}{{\tt :gr-qc/0102069}}].

\bibitem{qbounce6}
J.~Khoury, B.~Ovrut, P.~Steinhardt, and N.~Turok, {\it {Ekpyrotic universe:
  Colliding branes and the origin of the hot big bang}},  {\em Phys. Rev. D}
  {\bf 64} (2001) 123522, [\href{http://arxiv.org/abs/hep-th/0103239}{{\tt
  hep-th/0103239}}].

\bibitem{qbounce7}
F.~Alvarenga, J.~Fabris, N.~Lemos, and G.~Monerat, {\it {Quantum cosmological
  perfect fluid models}},  {\em Gen. Relat. Gravit.} {\bf 34} (2002) 651,
  [\href{http://arxiv.org/abs/gr-qc/0106051}{{\tt gr-qc/0106051}}].

\bibitem{qbounce8}
A.~Ashtekar, T.~Pawlowski, and P.~Singh, {\it {Quantum nature of the big
  bang}},  {\em Phys. Rev. Lett.} {\bf 14} (2006) 141301,
  [\href{http://arxiv.org/abs/gr-qc/0602086}{{\tt gr-qc/0602086}}].

\bibitem{qbounce9}
P.~Peter and N.~Pinto-Neto, {\it {Cosmology without inflation}},  {\em Phys.
  Rev. D} {\bf 78} (2008) 063506, [\href{http://arxiv.org/abs/0809.2022}{{\tt
  arXiv:0809.2022}}].

\bibitem{qbounce10}
A.~Ashtekar and P.~Singh, {\it {Loop quantum cosmology: A status report}},
  {\em Class. Quant. Grav.} {\bf 28} (2011) 213001,
  [\href{http://arxiv.org/abs/1108.0893}{{\tt arXiv:1108.0893}}].

\bibitem{qbounce11}
S.~Gielen and N.~Turok, {\it {Perfect quantum cosmological bounce}},  {\em
  Phys. Rev. Lett.} {\bf 117} (2016) 021301,
  [\href{http://arxiv.org/abs/1510.00699}{{\tt arXiv:1510.00699}}].

\bibitem{ob0}
A.~Ijjas and P.~Steinhardt, {\it {Implications of Planck2015 for inflationary,
  ekpyrotic and anamorphic bouncing cosmologies}},  {\em Class. Quantum Grav.}
  {\bf 33} (2016) 044001, [\href{http://arxiv.org/abs/1512.09010}{{\tt
  arXiv:1512.09010}}].

\bibitem{ob1}
Y.~Cai, F.~Duplessis, D.~Easson, and D.~Wang, {\it {Searching for a matter
  bounce cosmology with low redshift observations}},  {\em Phys. Rev. D} {\bf
  93} (2016) 043546, [\href{http://arxiv.org/abs/1512.08979}{{\tt
  arXiv:1512.08979}}].

\bibitem{ob2}
A.~Bacalhau, N.~Pinto-Neto, and S.~Vitenti, {\it {Consistent scalar and tensor
  perturbation power spectra in single fluid matter}},  {\em Phys. Rev. D} {\bf
  97} (2018) 083517, [\href{http://arxiv.org/abs/1706.08830}{{\tt
  arXiv:1706.08830}}].

\bibitem{ob3}
R.~Raveendran and L.~Sriramkumar, {\it {Viable scalar spectral tilt and
  tensor-to-scalar ratio in near-matter bounces}},  {\em Phys. Rev. D} {\bf
  100} (2019) 083523, [\href{http://arxiv.org/abs/1812.06803}{{\tt
  arXiv:1812.06803}}].

\bibitem{ob4}
I.~Agullo, J.~Olmedo, and V.~Sreenath, {\it {Predictions for the CMB from an
  anisotropic quantum bounce}},  {\em Phys. Rev. Lett.} {\bf 124} (2020)
  251301, [\href{http://arxiv.org/abs/2003.02304}{{\tt arXiv:2003.02304}}].

\bibitem{bounce-inflation}
P.~Singh, K.~Vandersloot, and G.~Vereshchagin, {\it {Non-Singular Bouncing
  Universes in Loop Quantum Cosmology}},  {\em Phys. Rev. D} {\bf 74} (2006)
  043510, [\href{http://arxiv.org/abs/gr-qc/0606032}{{\tt gr-qc/0606032}}].

\bibitem{Agullo:2020cvg}
I.~Agullo, D.~Kranas, and V.~Sreenath, {\it {Large scale anomalies in the CMB
  and non-Gaussianity in bouncing cosmologies}},  {\em Class. Quant. Grav.}
  {\bf 38} (2021), no.~6 065010, [\href{http://arxiv.org/abs/2006.09605}{{\tt
  arXiv:2006.09605}}].

\bibitem{PhysRevD.97.066021}
I.~Agullo, B.~Bolliet, and V.~Sreenath, {\it {Non-Gaussianity in loop quantum
  cosmology}},  {\em Phys. Rev. D} {\bf 97} (2018), no.~6 066021,
  [\href{http://arxiv.org/abs/1712.08148}{{\tt arXiv:1712.08148}}].

\bibitem{anomalies}
D.~Schwarz, C.~J. Copi, D.~Huterer, and G.~Starkman, {\it {CMB Anomalies after
  Planck}},  {\em Class. Quant. Grav.} {\bf 33} (2016), no.~18 184001,
  [\href{http://arxiv.org/abs/1510.07929}{{\tt arXiv:1510.07929}}].

\bibitem{Planck-nongaussian}
P.~C.~Y. Akrami and et~al., {\it {Planck 2018 results. IX. Constraints on
  primordial non-Gaussianity}},  {\em Astron. Astrophys.} {\bf 641} (2018) A9,
  [\href{http://arxiv.org/abs/1905.05697}{{\tt arXiv:1905.05697}}].

\bibitem{Planck:2018nkj}
{\bf Planck} Collaboration, N.~Aghanim et~al., {\it {Planck 2018 results. I.
  Overview and the cosmological legacy of Planck}},  {\em Astron. Astrophys.}
  {\bf 641} (2020) A1, [\href{http://arxiv.org/abs/1807.06205}{{\tt
  arXiv:1807.06205}}].

\bibitem{Planck:2018vyg}
{\bf Planck} Collaboration, N.~Aghanim et~al., {\it {Planck 2018 results. VI.
  Cosmological parameters}},  {\em Astron. Astrophys.} {\bf 641} (2020) A6,
  [\href{http://arxiv.org/abs/1807.06209}{{\tt arXiv:1807.06209}}].

\bibitem{Planck:2016aaa}
P.~A.~R. Ade, N.~Aghanim, M.~Arnaud, F.~Arroja, M.~Ashdown, J.~Aumont,
  C.~Baccigalupi, M.~Ballardini, A.~J. Banday, and et~al., {\it {Planck 2015
  results. XVII. Constraints on primordial non-Gaussianity}},  {\em Astronomy
  \& Astrophysics} {\bf 594} (Sep, 2016) A17.

\bibitem{Durrer:2020fza}
R.~Durrer, {\em {The Cosmic Microwave Background}}.
\newblock Cambridge University Press, 12, 2020.

\bibitem{DiDio:2018unb}
E.~Di~Dio, R.~Durrer, R.~Maartens, F.~Montanari, and O.~Umeh, {\it {The
  Full-Sky Angular Bispectrum in Redshift Space}},  {\em JCAP} {\bf 04} (2019)
  053, [\href{http://arxiv.org/abs/1812.09297}{{\tt arXiv:1812.09297}}].

\bibitem{Lesgourgues:2011re}
J.~Lesgourgues, {\it {The Cosmic Linear Anisotropy Solving System (CLASS) I:
  Overview}},  \href{http://arxiv.org/abs/1104.2932}{{\tt arXiv:1104.2932}}.

\bibitem{Blas:2011rf}
D.~Blas, J.~Lesgourgues, and T.~Tram, {\it {The Cosmic Linear Anisotropy
  Solving System (CLASS) II: Approximation schemes}},  {\em JCAP} {\bf 07}
  (2011) 034, [\href{http://arxiv.org/abs/1104.2933}{{\tt arXiv:1104.2933}}].

\bibitem{Planck:2015mym}
{\bf Planck} Collaboration, P.~A.~R. Ade et~al., {\it {Planck 2015 results. XV.
  Gravitational lensing}},  {\em Astron. Astrophys.} {\bf 594} (2016) A15,
  [\href{http://arxiv.org/abs/1502.01591}{{\tt arXiv:1502.01591}}].

\bibitem{Limber:1953}
D.~N. {Limber}, {\it {The Analysis of Counts of the Extragalactic Nebulae in
  Terms of a Fluctuating Density Field.}},  {\em Astrophys. J.} {\bf 117}
  (Jan., 1953) 134.

\end{thebibliography}\endgroup
\bibliographystyle{JHEP}

\end{document}